# Evolution of Resistive Switching Characteristics in WO$_{3-x}$-based MIM Devices by Tailoring Oxygen Deficiency

*Krishna Rudrapal$^a$, Biswajit Jana$^b$, Venimadhav Adyam$^c$ and Ayan Roy Chaudhuri$^{a,b*}$*

$^a$Advanced Technology Development Centre, Indian Institute of Technology Kharagpur, 721302, West Bengal, India

$^b$Materials Science Centre, Indian Institute of Technology Kharagpur, 721302, West Bengal, India

$^c$Cryogenic Engineering Centre, Indian Institute of Technology Kharagpur, 721302, West Bengal, India

ABSTRACT

We report on resistive switching (RS) characteristics of W/WO$_{3-x}$/Pt-based thin film memristors modulated by precisely controlled oxygen non-stoichiometry. RS properties of the devices with varied oxygen vacancy (V$_O$) concentration have been studied by measuring their DC current voltage properties. Switchability of the resistance states in the memristors have been found to depend strongly on the V$_O$s concentration in the WO$_{3-x}$ layer. Depending on x, the memristors exhibited forming-free bipolar, forming-required bipolar and non-formable characteristics. Devices with high V$_O$s concentration (~1×10$^{21}$ cm$^{-3}$) exhibited lower initial resistance and memory window of only 15, which has been increased to ~6500 with reducing V$_O$s concentration to ~5.8×10$^{20}$ cm$^{-3}$. Forming-free, stable RS with memory window of ~2000 have been realized for a memristor possessing V$_O$s concentration of ~6.2×10$^{20}$ cm$^{-3}$. Investigation of the conduction mechanism suggests that tailoring V$_O$s concentration modifies the formation and dimension of the conducting filaments as well as the Schottky barrier height at WO$_{3-x}$/Pt interface which deterministically modulates RS characteristics of the WO$_{3-x}$ based memristors.

INTRODUCTION

Resistive random access memory (RRAM), that stores data in the form of its resistance states defined by the history of applied voltage, has drawn significant research attention for next generation high dense non-volatile memory applications.[1] RRAMs which consist of simple metal-insulator-metal (MIM) structures offer various advantages, such as low power consumption, multiscale switching, scalability, fast operation speed and superior cycling endurance in comparison to the present day charge-based memory technologies. While resistive switching (RS) phenomenon has been reported in diverse class of materials, metal oxide-based RRAMs that involve modulation of resistance states due to change in the metal valence state (VCM) has drawn significant attention.[2], [3]

Various metal-oxides encompassing binary transition metal oxides (TMOs) such as $Ta_2O_5$, $TiO_2$, $WO_{3-x}$, $ZrO_2$, NiO, etc., and perovskites e.g. $Pr_{0.7}Ca_{0.3}MnO_3$, $SrTiO_3$, $SrZrO_3$, $BiFeO_3$ etc. have been investigated for their promising VCM type RS.[4]–[10] VCM type RS in metal-oxides often involves formation and rupture of conducting filaments (CFs) due to migration of oxygen vacancies ($V_{OS}$) under an applied electric field, their generation and annihilation at an active metal-oxide/electrode interface.[11] From the materials point of view, binary TMOs have been widely studied, which are attractive due to their simple structure, low cost, ease of processing, complementary metal–oxide–semiconductor (CMOS) technology compatibility, and promising performance.

Controlling the oxygen non-stoichiometry in binary TMOs through various approaches, such as formation of bilayer structures, utilization of different thin film growth techniques, modification of thin film growth conditions, incorporation of dopants, etc. have been found to play crucial roles in governing their RS properties. Sharath *et al.* demonstrated strong reduction of forming voltage in sub-stoichiometric $Ta_2O_{5-x}/TaO_x$ bilayer thin films with increasing $V_{OS}$ concentration in the $TaO_x$ layer grown under controlled oxygen atmosphere.[12] While bipolar RS had been observed for samples with higher $V_{OS}$ content, the switching transformed to threshold type for samples fabricated at the highest oxidation condition. Yong *et al.* compared the impact of different thin film growth techniques on $V_{OS}$ concentration in $HfO_x$ layers, which influences their forming voltage, memory window ($R_{off}/R_{on}$) and the Schottky barrier height (SBH) at $HfO_x$/TiN interface.[13] Skaja *et al.* reported manifestation of reduced forming voltage and high $R_{off}/R_{on}$ upon increasing $V_{OS}$ content in sputter deposited $Ta_2O_{5-x}$ thin films in which $V_{OS}$ concentration had been controlled by sputter power and $O_2$/Ar ratio.[14] Palhares *et al.* reported on Zr doping in $TaO_x$ thin films as an efficient method of reducing $V_{OS}$ formation energy, increasing confinement of the CFs which led to reduced device to device variability.[15] Ghenzi *et al.* demonstrated that small variation of $O_2$/Ar ratio during reactive sputter growth of $TiO_{2-x}$-based devices led to different RS characteristics, e.g. standard bipolar, electroforming free bipolar and non-switchable device within the same materials system owing to significant variation of $V_{OS}$ concentration.[5] Rehman *et al.* reported tuning of RS characteristics in Zn-doped $CeO_2$, where concentration and mobility of $V_{OS}$ in $CeO_2$ has been tuned by Zn dopant concentration.[16] Tunable digital to analogue RS in $Nb_2O_5$ thin films by $V_O$ engineering has been demonstrated by Xu *et al.*[17] Bousoulas *et al.* discussed that concentration and distribution of $V_{OS}$ in $TiO_{2-x}$ directly influence diameter of the CFs that impacts sensitivity of the conducting paths.[18]

Out of various binary TMOs which exhibit stable RS, $WO_3$, which is a n-type wide band gap TMO, has been a convenient choice for RS based device applications due to its compatibility with back-end-of-line processing in CMOS integrated circuits.[19] RS properties of $WO_3$ has been reported to be susceptible to oxygen non-stoichiometry in the oxide layer. Won *et al*. and Rudrapal *et al.* discussed the impact of $V_{OS}$ accumulation on the $WO_{3-x}$/electrode interface on determining its RS properties.[20], [21] Kim *et al.* reported the role of electrodes with different electronegativities in tuning RS characteristics of $WO_3$ layers by means of an interfacial suboxide layer formation.[22] Biju *et al.* discussed inhomogeneous distribution of $V_{OS}$ driven changeover of RS mechanism with changing thickness in thermally grown $WO_{3-x}$ films.[23] Yang *et al.* demonstrated on-demand realization of electrical and neuromorphic multifunction in $WO_{3-x}$-based nanoionic device through externally induced local migration of oxygen ions.[24] Rudrapal *et al.* demonstrated manifestation of forming-free (FF), self-compliant, multi-level RS in $WO_{3-x}$ layers, which has been attributed to reversible oxygen migration at $WO_{3-x}$/Pt interface.[25], [26] The existing literature reports suggest that, defect engineering based on tuning $V_{OS}$ concentration and assembly in $WO_{3-x}$ thin films holds the potential of deterministically control their RS properties. However, to the best of our knowledge report on the evolution of RS properties in $WO_{3-x}$ thin films with precise variation of $V_{OS}$ concentration is not present in the literature.

Here, we investigated precisely oxygen tailored $WO_{3-x}$ thin films and correlated variation of their RS characteristics with oxygen content in the layers. On the basis of $V_{OS}$ concentration modulation in $WO_{3-x}$, we highlighted an effective way to control operating parameters such as forming requirement, set voltage ($V_{set}$), and $R_{off}/R_{on}$ in W/$WO_{3-x}$/Pt asymmetric MIM devices and discussed the corresponding switching mechanism. This study enables development of fabrication strategies based on $V_{OS}$ tailoring for realizing CMOS compatible $WO_{3-x}$ based forming free bipolar RS memory devices with a high $R_{off}/R_{on}$ ratio.

RESULTS AND DISCUSSION

All the $WO_{3-x}$ films deposited at room temperature are amorphous in nature which has been confirmed from grazing incidence X-ray difraction (GIXRD) measurements (Figure S1). The samples under investigation possess similar thickness (~42-47 nm), low roughness (<1 nm) and sharp interfaces with the electrodes without the formation of any interfacial layers (Figures S2 and S3).

Figure 1 compares high-resolution X-ray photoelecron spectrocopy (XPS) data around the W 4f and O 1s region for S75, S80, and S85. The binding energies of 35.3 eV and 37.5 eV corresponding to $W^{6+}$ $4f_{7/2}$ and $W^{6+}$ $4f_{5/2}$ respectively.[27] Additionally, two minor peaks appeared at 34.3 eV and 36.5 eV corresponding to the $W^{5+}$ $4f_{7/2}$ and $W^{5+}$ $4f_{5/2}$ respectively. The $4f_{5/2}$ and $4f_{7/2}$ peaks are separated by an expected spin-orbit splitting of 2.2 eV and an area ratio

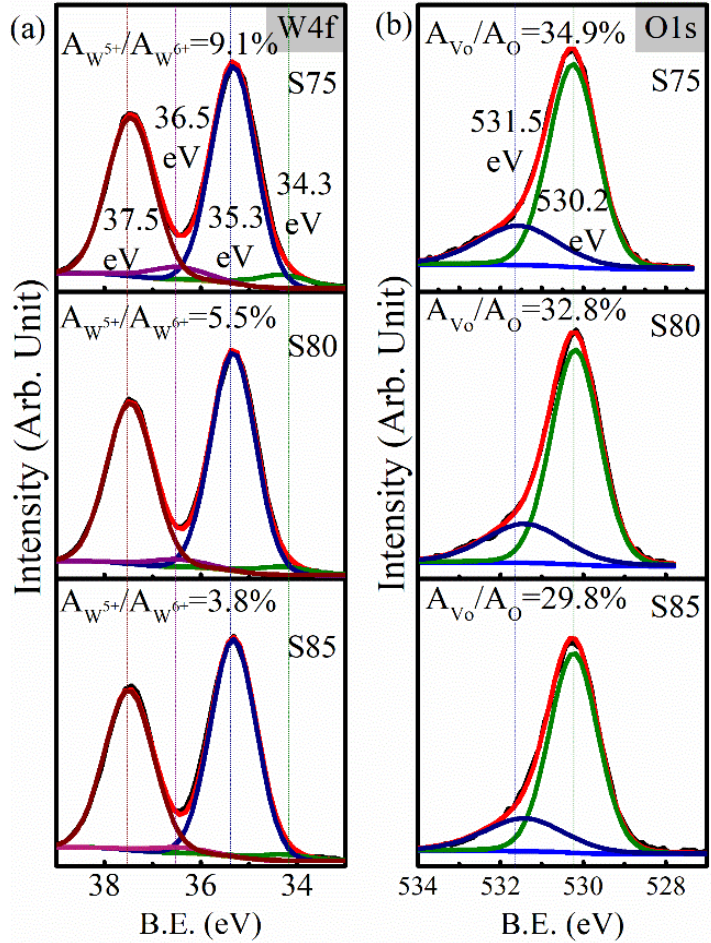

**Figure 1.** High resolution XPS spectra from sample S75, S80, and S85 with its deconvoluted peak and corresponding area ratio representing variation of stoichiometry with $O_2$ partial pressure during the growth. (a) W 4f spectra and its corresponding area ratio of $W^{5+}$ to $W^{6+}$ peaks, (b) O 1s and its corresponding area ratio of its non-stoichiometric to lattice oxygen peaks.

of 4:3 for both the oxidation states of W.[20] Manifestation of $W^{5+}$ states in the spectra indicates the formation of non-stoichiometric oxide layers.[27] In case of $WO_3$, the presence of $W^{5+}$ species is commonly associated with the formation of $V_O$s to maintain charge neutrality.[20], [28] Oxygen non-stoichiometry in the $WO_{3-x}$ layers has also been confirmed from the O 1s peak analysis. The O 1s peak (Figure 1b) could be deconvoluted into two sub-peaks at 530.2 eV and 531.5 eV respectively. While the peak at 530.2 eV represents the oxygen at lattice positions of the stoichiometric oxide layer, the higher energy peak at 531.5 eV represents O atoms in the oxygen-deficient regions of the $WO_{3-x}$ matrix.[25] The degree of

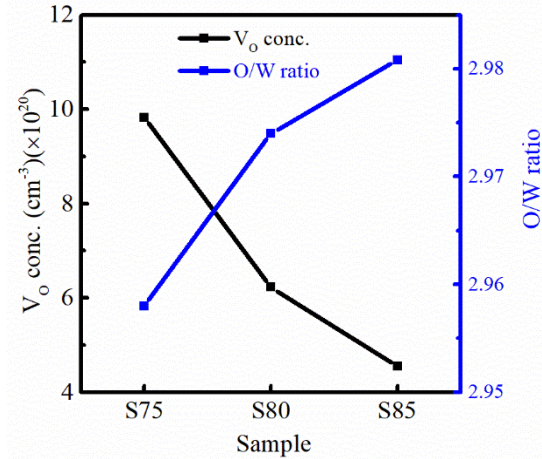

**Figure 2.** Calculated $V_O$ concentration and O/W ratio for samples S75, S80, and S85. Lines are to guide the eyes. Decrease of $V_O$ and increase of O/W ratio from S75 to S85 is clearly evident.

oxygen non-stoichiometry in the $WO_{3-x}$ layers can be correlated to the area ratio of the O 1s peaks corresponding to the oxygen deficiency and stoichiometric oxygen ($A_{Vo}/A_O$). The $A_{Vo}/A_O$ ratio has been found to reduce monotonically from 34.9% for S75 to 29.8% for S85. This indicates improvement of oxygen stoichiometry in the sputter-grown $WO_{3-x}$ layers with increasing $O_2$ partial pressure during deposition. The amount of $V_{OS}$ in the $WO_{3-x}$ layer has been estimated from the $W^{5+}:W^{6+}$ area ratio ($A_W^{5+}/A_W^{6+}$).[21] The $A_W^{5+}/A_W^{6+}$ has been found to reduce from 9.1% for S75 to 3.8% for S85. This corresponds to a monotonous reduction of $V_{OS}$ concentration in the $WO_{3-x}$ layers from $1\times10^{21}$ cm$^{-3}$ (S75) to $4.5\times10^{20}$ cm$^{-3}$ (S85) (Figure 2), representing a change of O/W ratio from ~2.95-2.98. A similar trend of oxygen partial pressure-dependent evolution of $V_{OS}$ content in DC reactive sputter growth metal oxide thin films has been discussed by Skaja *et al.* for $Ta_2O_{5-x}$ and Ghenzi *et al.* for $TiO_{2-x}$.[5], [14]

RS characteristics of the non-stoichiometric $WO_{3-x}$ layer have been investigated using $W/WO_{3-x}/Pt$ asymmetric MIM structure. The $WO_{3-x}$ layer possesses sharp interfaces with both top and bottom electrodes as evidenced from cross sectional transmission electron microscopy (TEM) image of a representative sample presented in Figure S3. The inset of Figure 3a represents a schematic of the device structure where DC bias has been applied to the top W electrode and the bottom Pt electrode has been kept grounded. Prior to the RS measurements, initial resistance ($R_i$) of the as-fabricated devices has been measured at 0.1 V. Figure 3a represents $R_i$ of all the samples including the device-to-device variability measured over 15 devices from each of the samples. The median value of $R_i$ has been found to monotonically increase from 13 MΩ for S75 to 32 GΩ for S85. Such large variation of intrinsic resistance level by more than three orders of magnitude has been ascribed to difference in concentration of $V_{OS}$ in the $WO_{3-x}$ layers. Lowering of their resistance with increasing $V_{OS}$ is related to the increase in free electron concentration, as $V_{OS}$ act as donors in $WO_3$.[29] Considering that $V_{OS}$ in $WO_{3-x}$ predominantly manifest as $V_O^{2+}$, excess electron concentration in the layers due to $V_O$ formation varies between ~$2\times10^{21}$ cm$^{-3}$ (S75) to ~$9\times10^{20}$ cm$^{-3}$ (S85).[29] The increase in n-type characteristics of the $WO_{3-x}$ layers with increasing $V_{OS}$ concentration has also been confirmed from the reduced separation of Fermi level from the conduction band edge (~1.3 eV for S85 to ~0.7 eV for S75) estimated from investigation of XPS valence band spectra and absorbance spectra from UV-Vis diffuse reflectance spectroscopy (DRS) measurements (Figures S4 and S5).

Figure 3b-f represents room temperature I-V characteristics of the samples under DC voltage sweeps of more than 100 cycles measured in the sequence of 0.0 V→-3.0 V→0.0 V→+3.0 V→0.0 V with a step of 0.05 V and a compliance current set at 1 mA. The compliance current and voltage range used during DC sweep measurements are known to influence on the RS characteristics of the MIM devices significantly.[30] Therefore, the RS characteristics of the samples S75-S83 under investigation have been compared using a constant compliance current (1 mA) and DC bias range (± 3V). Initially, all the as prepared devices have been at high resistance state (HRS) and all, but S85 turned into low resistance state (LRS) when a suitable voltage with negative polarity ($V_{set}$) has been applied on the top W electrodes initiating the set process. The samples have been reset back to HRS on application of a positive DC bias, representing a bipolar clockwise switching (CWS). Interestingly, the samples S75-S80 having

$V_{OS}$ concentration in the range of ~1 ×10$^{21}$ cm$^{-3}$ - 6.2 ×10$^{20}$ cm$^{-3}$ exhibited stable bipolar switching without the requirement of any additional electroforming step. This FF nature could be attributed to the large $V_{OS}$ concentration in the WO$_{3-x}$ layers. FF bipolar RS in non-stoichiometric WO$_{3-x}$ having similar $V_{OS}$ density has been reported earlier.[25] However, the sample S83 required an additional forming step that involves application of a larger negative DC bias (~-4.15 V) prior to exhibiting RS (Figure 3e). Thus, S83 which possesses a lower concentration of $V_{OS}$ (~5.8 ×10$^{20}$ cm$^{-3}$) has been termed as a forming-required (FR) device. Further reduction of $V_{OS}$ concentration to ~4.5 ×10$^{20}$ cm$^{-3}$ (S85) rendered the layer to be highly insulating, which could not be switched to LRS even after applying a DC voltage as high as -10 V (inset of Figure 3f), indicating a non-formable (NF) device. This has been attributed to insufficient $V_{OS}$ density in the WO$_{3-x}$ layer which hinders the formation of CFs in the oxide layer.[5] Variation of the I-V characteristics in the WO$_{3-x}$ samples with controlled non-stoichiometry clearly exhibits the deterministic role of $V_{OS}$ on their bipolar RS properties, which change from FF to FR to finally NF device depending on the $V_{OS}$ concentration. $V_{OS}$ concentration-dependent modulation of RS characteristics has been reported earlier in literature for various metal oxides. Sharath *et al.* reported $V_{OS}$ concentration-dependent modulation from FR to FF RS characteristics in HfO$_{2-x}$ thin films.[31] Skaja *et al.* reported a reduction of forming voltage in the case of Ta$_2$O$_{5-x}$ with the increase of $V_{OS}$ concentration.[14] Ghenzi *et al.* observed NF device when the oxygen partial pressure during the TiO$_{2-x}$ layer growth has been increased beyond a certain limit.[5]

In the present study, the nature of RS in all the WO$_{3-x}$ samples under investigation has been associated with gradual change in current with DC voltage sweep. In this case, the V$_{set}$ has been defined to be the applied voltage at which current reaches the compliance value, or the maximum applied voltage (-3V) in case current remains below the compliance limit. V$_{set}$ determined for the samples S75-S83 based on this convention has been plotted in Figure 4a. The median value of V$_{set}$ has been found to increase from -1.9 V to -3 V for samples S75 to S83 respectively. Monotonic increase of V$_{set}$ from S75 to S83 is related to the reduced concentration of V$_{OS}$ in the WO$_{3-x}$ layer. Park *et al.* reported an increased V$_{set}$ when the V$_{OS}$ concentration has been reduced by increasing the O$_2$ flow during the growth of TiO$_x$ layer.[32] It is to be noted that for all the samples the reset voltage (V$_{reset}$) has been fixed at 3 V.

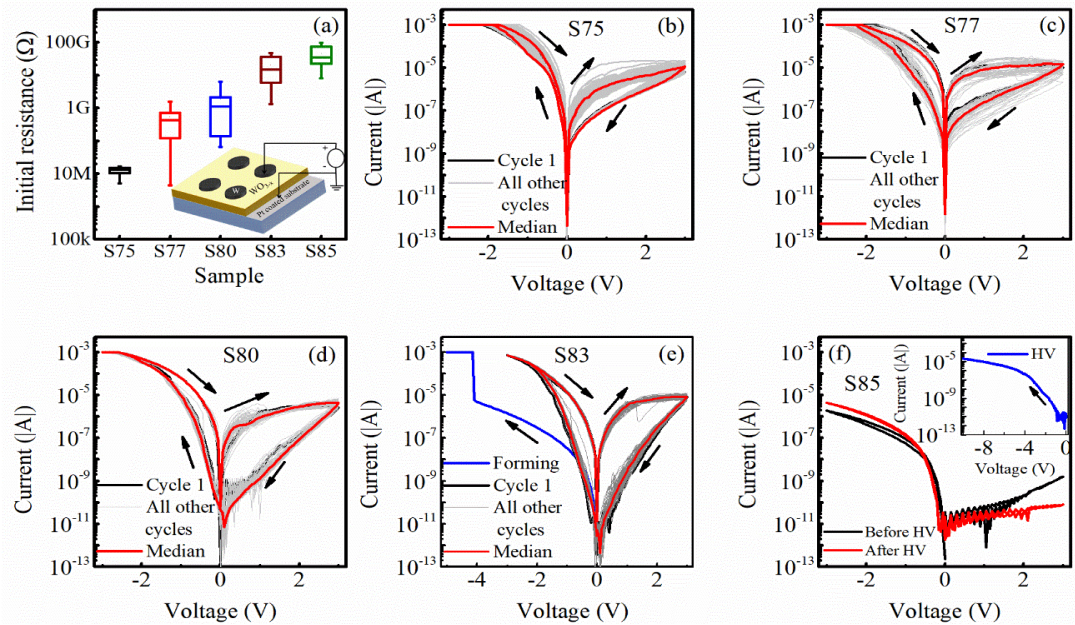

**Figure 3.** (a) Initial resistance from all the samples taken from 15 devices for each case, inset shows the schematic of the W/WO$_{3-x}$/Pt device structure with its electrical connections. DC sweep characteristics of the samples with more than 100 sweep cycles (b) S75, (c) S77, (d) S80, (e) S83. (f) DC sweep cycles before and after an application of HV step shows the S85 is a non-switchable device, inset shows the applied HV step. The figure represents evolution of RS with varying V$_O$-concentrated sample.

Figure 4b displays $R_{off}/R_{on}$ of the $WO_{3-x}$ samples determined from the median switching cycles at a read voltage ($V_{read}$) of +0.5 V. $R_{off}/R_{on}$ increased significantly from ~15 for S75 to ~6500 for S83 indicating an inverse relation with the $V_O$s concertation in the switching layer.

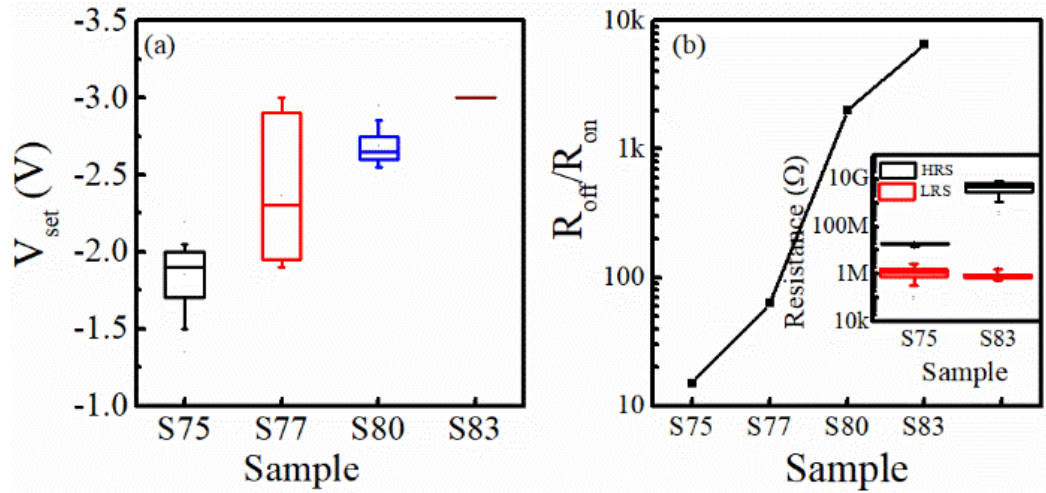

**Figure 4.** Variation of RS parameters among the samples (a) distribution of $V_{set}$ for each switched sample, it increased from S75 to S83. (b) Extracted $R_{off}/R_{on}$ ratio, it increased almost exponentially from S75 to S83, lines are to guide the eyes, the inset shows the variation of HRS and LRS of S75 and S83.

In order to elucidate the observed trend of $R_{off}/R_{on}$ further, LRS and HRS resistance values (at +0.5 V) of the end members (S75 and S83) of the set of samples considered in this study, has been plotted for 100 cycles and 5 devices (inset of Figure 4b). While the LRS varied marginally around 1 MΩ, the HRS increased significantly (more than three orders of magnitude) from S75 to S83. Clearly, higher HRS resistance in samples with lower $V_O$s concentration leads to the enhanced $R_{off}/R_{on}$. This has been be attributed to the inferior insulating characteristic of the $V_O$ rich samples and reduced oxygen available to rupture the filament.[31], [33]

SWITCHING MECHANISM

The switching mechanism has been elucidated by investigating the conduction mechanism before and after the switching event. For that, the temperature-dependent DC sweeps were taken for samples from S75 to S83 and the corresponding LRS and HRS data have been plotted against temperature as shown in Figure 5a and b respectively. For all the $WO_{3-x}$ samples under investigation exhibiting RS, resistance at LRS has been found to increase with increasing temperature, whereas in HRS an opposite trend has been observed. This indicates formation of metallic filaments during the set operation which predominantly control the LRS current, whereas rapture of filaments during the reset step forms the HRS in which the current is controlled by the $WO_{3-x}$/Pt Schottky junction.[25]

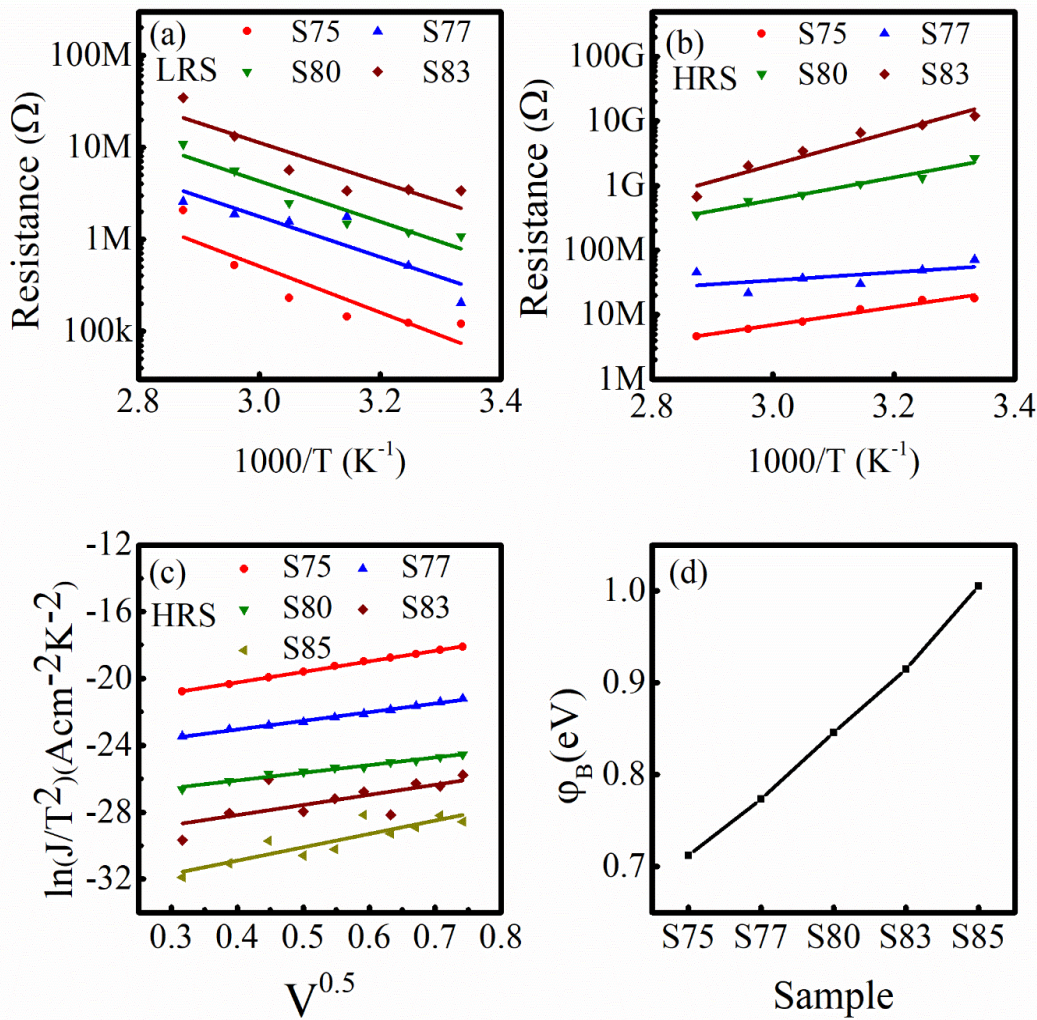

**Figure 5.** Investigation of the conduction mechanism in the LRS and HRS state. Change of resistances with temperature for all the switchable samples at (a) LRS, indicating metallic-like behavior. (b) HRS, indicating Schottky-like behavior. (c) Fitting of the HRS data for the Schottky emission. (d) Extracted $\varphi_B$ from each of the sample which shows a monotonic increment from S75 to S85, lines are to aid the eyes.

In order to investigate possible dependence of the Schottky barrier at WO$_{3-x}$/Pt junction on V$_{OS}$ concentration in the oxide layer, current conduction in the HRS has been analyzed under the framework Schottky emission theory: [34]

$$J = A^* T^2 e^{-\frac{q\left(\varphi_B - \sqrt{\frac{qV}{4\pi\varepsilon_r\varepsilon_0 d}}\right)}{k_B T}} \quad (1)$$

Where $J$ is current density, $A$ is the effective Richardson constant, $T$ is absolute temperature, $q$ is the electronic charge, $\varphi_B$ is SBH, $V$ is voltage, $\varepsilon_r, \varepsilon_0$ are the relative and absolute permittivity respectively, $d$ is the space charge width, $k_B$ is the Boltzmann constant.

Figure 5c represents the Richardson's plot [ln($J/T^2$) vs V$^{0.5}$] for the Schottky conduction at HRS. The linear relationship of ln($J/T^2$) vs V$^{0.5}$ is consistent with the Schottky conduction mechanism. Figure 5d demonstrates evolution of SBH ($\varphi_B$) at the WO$_{3-x}$/Pt interface with changing V$_{OS}$ concentration in the sub-stoichiometric WO$_{3-x}$ layers. $\varphi_B$ has been found to decrease from 1 eV for S85 to 0.71 for S75, which is consistent with the reduction of SBH with increasing V$_{OS}$ in metal oxide thin films.[35]–[38]

Next, microscopic mechanism of RS phenomenon in the asymmetric devices with WO$_{3-x}$ layers of varying stoichiometry has been elucidated considering metallic filament-driven LRS and Schottky barrier-driven HRS current. Figure 6 presents a schematic of the proposed RS mechanism in the WO$_{3-x}$ layers considering the S75 and S83 as representative V$_{OS}$ rich and V$_O$ poor samples respectively. It has been postulated that the V$_O$ concentration is higher at the WO$_{3-x}$/Pt interface compared to the other parts of the layer. Such a postulation is valid considering that Pt is permeable to oxygen which allows reversible migration and storage of oxygen species in the grain boundaries of Pt in contact with metal oxides.[39]–[41] RS in the present samples involves V$_O$ diffusion due to the concentration gradient, their drift in response to the applied potential and oxygen transfer reaction at the interface with the inert Pt electrode.[25] When a negative bias is applied on the top W electrode, conductive filament consisting of relatively large density of $V_{OS}$ starts to grow from the cathode toward the anode due to the drift and diffusion flux of $V_{OS}$. Additionally, $V_{OS}$ gets generated by anodic oxidation at the $WO_{3-x}$/Pt interface:[35]

$$V_O^{\bullet\bullet}(Pt) + O_O^\times(WO_{3-x}) \rightarrow O_O^\times(Pt) + V_O^{\bullet\bullet}(WO_{3-x}) \qquad (2)$$

A gradual increase of current with applied bias in all the W/$WO_{3-x}$/Pt devices under investigation (Figure 3) suggests a competition between $V_O$ drift and diffusion migration towards W cathode, and oxygen exchange reaction at the high-work function Pt electrode. Considering a much lower diffusion coefficient of oxygen in Pt compared to $WO_3$, certain depletion of $V_{OS}$ in $WO_{3-x}$ near its interface with Pt can take place.[42], [43] In such case, $V_{OS}$ migration towards cathode shall dominate over the oxygen extraction at anode, giving rise to a high insulating region near the $WO_{3-x}$/Pt interface. Formation of a $V_O$ depleted region near the anode induces a reverse concentration gradient (from cathode to anode) which eventually drives $V_{OS}$ diffusion towards the anode and reduces the depletion gap. At an adequately large applied voltage, the $V_{OS}$ generation flux at the anode also increases which aids to the filament formation. Finally at the $V_{set}$, the formation of the conductive filament gets completed when LRS sets in. The observed difference in $V_{set}$ for all the samples under consideration has been attributed to their initial $V_{OS}$ concentration and $V_{OS}$ generation at the $WO_{3-x}$/Pt interface. As evident from Figure 3a, electrical conductivity (σ) of the $WO_{3-x}$ layers increases with increase in $V_{OS}$ concentration in the layer. Considering $V_{OS}$ formation barrier at a metal-oxide/anode interface to be a constant, the $V_{OS}$ depletion gap varies inversely with σ of the oxide layer.[44] Therefore, it is postulated that $V_{OS}$ depletion gap at the $WO_{3-x}$/Pt interface increases from S75 to S85 in accordance to their increasing electrical resistances. Since $V_{OS}$ generation rate at the oxide/anode interface gets prompted by the heat generation due to Joule heating effect, $V_{set}$ reduces from S83 (having lower σ) to S75 (having higher σ). This is further established by the fact that while $WO_{3-x}$ samples S75-S80 ($[V_O] \sim 1 \times 10^{21}$ cm$^{-3}$-$6.2 \times 10^{20}$ cm$^{-3}$) exhibit forming free RS, a forming step is required for the sample S83 ($[V_O] \sim 5.8 \times 10^{20}$ cm$^{-3}$) and S85 ($[V_O] \sim 4.5 \times 10^{20}$ cm$^{-3}$) did not exhibit any RS at all even after applying a voltage as high as -10 V.

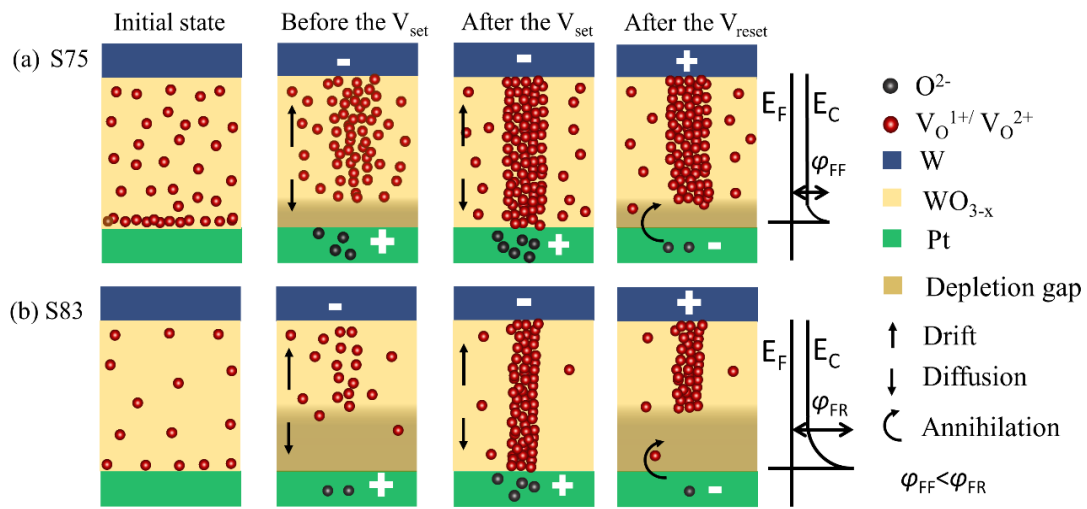

**Figure 6.** Schematic representation of the proposed switching mechanism through the initial, set, and reset process along with the evolution of filament and depletion gap depending on the $V_{OS}$ concentration for (a) S75, (b) S83.

In order to further elucidate the influence of CFs on the conduction behaviour of $WO_{3-x}$ layers, resistance states of S75 and S83 have been compared at -0.5V (Figure S6). While resistance of the LRS increased by 10 times from S75 to S83, the HRS increased by 1000 times. Under the framework of filamentary conduction model, increase LRS resistance with decreasing $V_{OS}$ suggests a higher filament resistance. This is attributed to the reduction of filament diameter with reducing $V_{OS}$ concentration from S75 to S83. Reduction of CFs' diameter with reducing $V_{OS}$ concentration in metal-oxide layers has been reported for $TiO_{2-x}$ and $NbO_x$ based RS devices.[5], [18], [45] Such difference in the LRS resistance has not been observed when measured at +0.5V during reset cycles (Figure 4b). During the reset operation for bipolar RS observed in the present study, when a positive bias is applied on the W top electrode, $V_{OS}$ annihilation at the $WO_{3-x}$/Pt takes place:

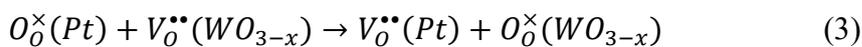
$$O_O^\times(Pt) + V_O^{\bullet\bullet}(WO_{3-x}) \rightarrow V_O^{\bullet\bullet}(Pt) + O_O^\times(WO_{3-x}) \qquad (3)$$

As a result, partial dissolution of filament starts to re-establish the Schottky junction, and current conduction during the reset cycle begins to get controlled also by the Schottky barrier at the $WO_{3-x}$/Pt interface even before complete reset. The nature of the $V_{OS}$ depleted region and the CF determines the LRS resistance during reset operation. Similarly, large difference of the HRS resistance between S75 and S83 at a read voltage of -0.5V is attributed to the better oxidation of the samples which present a higher barrier to the conduction of electrons prior to the set process.

CONCLUSIONS

In summary, fabrication of amorphous $WO_{3-x}$ thin films by DC reactive sputtering under varying the $O_2$/Ar ratio provides an effective method of inducing precisely controlled oxygen non-stoichiometry in the oxide layers. XPS investigation confirmed stoichiometry variation which corresponds to $V_{OS}$ concentration ranging between ~4.5 ×$10^{20}$ cm$^{-3}$ to ~1 ×$10^{21}$ cm$^{-3}$ in the $WO_{3-x}$ layers grown under different $O_2$/Ar ratio. Increase in $V_{OS}$ with reduced $O_2$/Ar ratio has been found to increase conductivity of the oxide layers which is attributed to the enhanced free electron concentration in the n-type $WO_{3-x}$. DC I-V measurements on the W/$WO_{3-x}$/Pt devices provided the evidence that $V_{OS}$ concentration plays a significant role in their RS characteristics. Low concentration of $V_{OS}$ (≤4.5×$10^{20}$ cm$^{-3}$) suppresses the formation of CFs thus rendering the devices non-switchable. Increase in $V_{OS}$ concentration to ~ 5.8×$10^{20}$ cm$^{-3}$ leads to a switchable device after inducing an electroforming step, while the devices with higher $V_{OS}$ concentration ($V_O$ = 6.2 ×$10^{20}$ cm$^{-3}$ to 1 ×$10^{21}$ cm$^{-3}$) exhibit forming free bipolar RS. Further, the set voltage and memory window of the devices have been found to reduce with increasing $V_{OS}$ concentration. Analysis of the DC I-V characteristics establishes that modulation of RS characteristics in the $WO_{3-x}$ layers relates to the CF diameter, their formation and dissolution due to $V_{OS}$ diffusion under concentration gradient, their drift in response to the applied potential and oxygen transfer reaction at the $WO_{3-x}$/Pt interface. Our work has provided strong proof that RS characteristics of CMOS compatible are deterministically tunable by $V_{OS}$ engineering.

EXPERIMENTAL DETAILS

The $WO_{3-x}$ layers have been deposited on Pt (~34 nm)/Ti(~5 nm)/$SiO_2$(~300 nm)/Si(100) substrate. Before any deposition, the substrates have been cleaned by dipping sequentially in Trichloroethylene, Acetone, and Methanol for 5 min for each case in an ultrasonic bath. After that, it has been rinsed with de-ionized (DI) water and dried with a nitrogen blower. The chamber had been evacuated to a base pressure of $7.5 \times 10^{-4}$ mTorr prior to the oxide layer deposition. The depositions were carried out at room temperature by DC reactive sputtering using a commercial tungsten target (advanced engineering materials, purity >99.99%) with a chamber pressure of 5 mTorr. In order to achieve varied oxygen non-stoichiometry in the $WO_{3-x}$ layers, five samples (Table 1) have been grown with oxygen partial pressure during the growth varying from 75% to 85% in a mixture of oxygen and argon gas ($O_2$/Ar) keeping all the other parameters unchanged. The samples have been prepared by maintaining a target-to-substrate distance of 10 cm and DC power density of 1.23 Watt/$cm^2$. MIM structures for investigating RS properties have been prepared by depositing W dots with a diameter of 100 μm on the $WO_{3-x}$ thin films by DC sputtering through a shadow mask.

The structural property has been investigated by GIXRD using X-ray source of CuKα1 having a wavelength of 1.5406 Å (PANalytical, Empyrean). The diffraction patterns were collected with a step size of 0.06° in the 2θ range of 20–60°. The incidence angle of CuKα radiation was kept at 0.4° relative to the sample surface. The X-ray reflectivity (XRR) measurements were done in the same instrument and the fittings were done using the AMaaS software. The device structure was confirmed by a cross-section transmission electron microscope (TEM) (FEI TECNAI G2 F20 X-TWIN) image. Bandgap was measured through UV-Vis DRS. Chemical analysis and valence band spectra of the oxide layers have been done using XPS (PHI 5000 VersaProbe III, USA) measurements with monochromatic Al Kα X-rays (spot diameter 100 μm). Deconvolution of the XPS spectra includes Shirley-type baseline subtraction and profile fitting through the Lorentzian-Gaussian function. The C1s peak at 284.5 eV has been taken as a reference point for all the binding energy calculations. DC current-voltage (I-V) measurements were carried out by Keithley 4200-SCS keeping a step size of 0.05 V. A JANIS probe station (model no: ST500-UHT-1-(4CX)) has been used to probe the samples.

Table 1. Variation of sputtering deposition parameters for $WO_{3-x}$ films

| Sample | $O_2$ partial pressure |
|---|---|
| S75 | 75% |
| S77 | 77% |
| S80 | 80% |
| S83 | 83% |
| S85 | 85% |

AUTHOR INFORMATION

Corresponding Author

*A. Roy Chaudhuri

Email: ayan@matsc.iitkgp.ac.in

Author Contributions

A.R.C conceived the idea. A.R.C and K.R designed the experiments, executed the measurements and analyzed the data. B.J contributed in the experiments. A.R.C and V.A supervised the work. All the authors discussed the results and have given approval to the final version of the manuscript.

Funding Sources

Authors acknowledge partial funding from SERB, SPARC scheme, The Ministry of Education, Govt. of India.

Notes

Authors declare no competing financial interest.

ACKNOWLEDGEMENT

Authors acknowledge SERB (Letter No CRG/2021/000811) and the Ministry of Education, Govt. of India (SPARC, vide letter No. SPARC/2018-2019/P252/SL); for partial financial support of the work, the Central Research Facility (CRF) of Indian Institute of Technology Kharagpur for various characterization facilities. Authors acknowledge Mr. Subhajit Dutta and Dr. BN Shivakiran Bhaktha, Department of Physics, IIT Kharagpur for UV-Vis DRS measurements.